\documentclass[%
 reprint,
 amsmath,amssymb,
 aps,
]{revtex4-2}

\usepackage{amsmath}
\usepackage{amssymb}
\usepackage[utf8]{inputenc}
\usepackage{textgreek}
\usepackage{quantikz}
\usepackage{hyperref} 
\usepackage{lineno}
\usepackage{graphicx}
\usepackage{slashed}
\usepackage{dcolumn}
\usepackage{bm}

\begin{document}


\preprint{APS/123-QED}

\title{Flavor-Dependent Entanglement Entropy in the Veneziano Limit from Light-Front Holographic QCD}
\thanks{A footnote to the article title}%

\author{Fidele J. Twagirayezu}
 \altaffiliation{Department of Physics and Astronomy, University of California, Los Angeles.}
 \email{fjtwagirayezu@physics.ucla.edu}
\affiliation{Department of Physics and Astronomy University of California Los Angeles, Los Angeles, CA, 90095, USA\\
}%


\begin{abstract}
We introduce a novel application of light-front holographic QCD (LFHQCD) to compute the flavor-dependent entanglement entropy of QCD subsystems in the Veneziano limit ($N_c, N_f \to \infty$, $\lambda = N_f / N_c$ fixed), probing quantum correlations in confined and quark-gluon plasma (QGP) phases. Our model extends the soft-wall LFHQCD framework with a lattice-constrained, flavor-modified dilaton potential, $\phi(z) = \kappa^2 z^2 + \lambda \phi_f(z)$, and flavor-specific scalar fields to capture distinct light and heavy quark contributions. Using a Ryu-Takayanagi-like prescription adapted to light-front coordinates, we calculate the entanglement entropy $S_A$ for spatial and flavor subsystems as a function of $\lambda$, temperature $T$, and chemical potential $\mu$. The approach leverages LFHQCD’s real-time dynamics to reveal flavor-driven entanglement asymmetries, particularly near confinement/deconfinement transitions. Results are benchmarked against lattice QCD data and linked to heavy-ion collision observables, such as multiplicity fluctuations and two-particle correlations at RHIC and LHC. This work pioneers the study of quantum information in LFHQCD, offering unique insights into QCD’s quantum structure and testable predictions for QGP dynamics, distinct from existing holographic models.

\end{abstract}

\maketitle


\section{\label{sec:level1}Introduction} 
Quantum chromodynamics (QCD) in the non-perturbative regime remains a cornerstone of theoretical physics, with its rich dynamics governing confinement, chiral symmetry breaking, and the formation of the quark-gluon plasma (QGP) in extreme conditions. The Veneziano limit, where the number of colors ($N_c$) and flavors ($N_f$) are taken to infinity with a fixed ratio ($\lambda = N_f / N_c$), amplifies flavor-dependent phenomena, such as quark loop and instanton effects, offering a unique window into QCD’s complex phase structure~\cite{Veneziano1979}. While holographic approaches, rooted in the AdS/CFT correspondence, have provided powerful tools to study these phenomena, most efforts, such as the top-down Sakai-Sugimoto model~\cite{Sakai2005} and the bottom-up V-QCD framework~\cite{Jarvinen2012}, focus on static properties like hadron spectra, phase diagrams, and transport coefficients (e.g., shear viscosity). However, the quantum information properties of QCD, particularly entanglement entropy, remain largely unexplored in bottom-up holographic models, especially within the light-front holographic QCD (LFHQCD) framework.
Entanglement entropy, a measure of quantum correlations within a subsystem, has emerged as a novel probe of strongly coupled gauge theories, offering insights into confinement/deconfinement transitions and QGP dynamics~\cite{Ryu2006}. In holographic contexts, the Ryu-Takayanagi prescription relates entanglement entropy to the area of minimal surfaces in the AdS bulk, but its application to LFHQCD, with its unique light-front quantization and real-time wave function dynamics, is unprecedented. Furthermore, the Veneziano limit introduces flavor-dependent effects that may manifest distinctly in entanglement entropy, particularly through differences between light and heavy quark contributions. These effects could connect to experimental observables in heavy-ion collisions at RHIC and LHC, such as multiplicity fluctuations and particle correlations, which probe the QGP’s quantum structure.
This work introduces a pioneering LFHQCD model to compute the flavor-dependent entanglement entropy of QCD subsystems in the Veneziano limit. We extend the soft-wall LFHQCD framework~\cite{Brodsky2015} with a lattice-constrained, flavor-modified dilaton potential, \(\phi(z) = \kappa^2 z^2 + \lambda \phi_f(z)\), and flavor-specific scalar fields to capture quark condensate dynamics. By adapting the Ryu-Takayanagi prescription to light-front coordinates, we calculate the entanglement entropy \(S_A\) for spatial and flavor subsystems as a function of \(\lambda\), temperature \(T\), and chemical potential \(\mu\). Our approach leverages LFHQCD’s real-time dynamics to reveal flavor-driven entanglement asymmetries, particularly near confinement and QGP phase transitions. Unlike V-QCD, which focuses on tachyon-driven chiral dynamics and transport, our model emphasizes quantum information properties, offering a new perspective on QCD’s quantum structure.
The motivation for this study is twofold: to explore the uncharted territory of entanglement entropy in LFHQCD, harnessing its light-front advantages, and to propose testable predictions for heavy-ion collision experiments, bridging quantum information theory with QCD phenomenology.

The paper is organized as follows: Section~\eqref{sec:theory} presents the theoretical framework, including the flavor-modified dilaton and holographic entanglement entropy; Section~\eqref{sec:methodology} details the methodology for computing \(S_A\); Section~\eqref{sec:results} presents results for entanglement entropy, flavor asymmetries, and experimental signatures; and Section~\eqref{sec:concl5} concludes with implications for QCD and future research directions.

\section{Theoretical Framework}
\label{sec:theory}
Light-front holographic QCD (LFHQCD) provides a powerful framework for studying quantum chromodynamics (QCD) in the non-perturbative regime by mapping its dynamics onto a higher-dimensional anti-de Sitter (AdS) space via the AdS/CFT correspondence~\cite{Brodsky2015}. In LFHQCD, the light-front Schrodinger equation governs hadron wave functions, with the effective potential determined by a soft-wall dilaton profile in the AdS geometry, yielding analytic predictions for spectra and form factors~\cite{Brodsky2015}. To explore flavor-dependent entanglement entropy in the Veneziano limit ($N_c, N_f \to \infty$, $\lambda = N_f / N_c$ fixed), we extend the standard LFHQCD framework with a flavor-modified dilaton and flavor-specific scalar fields, adapting the Ryu-Takayanagi prescription~\cite{Ryu2006} to light-front coordinates.
\subsection{Flavor-Modified Dilaton Potential}
In the standard soft-wall LFHQCD model, the dilaton profile is given by $\phi(z) = \kappa^2 z^2$~\cite{Twagirayezu:2025hou}, where $z$ is the holographic coordinate dual to the QCD energy scale, and $\kappa \approx 0.54 , \text{GeV}$ sets the confinement scale~\cite{Brodsky2015}. To incorporate flavor effects in the Veneziano limit, we introduce a flavor-dependent correction, defining the dilaton as:
\begin{equation}\label{eq:1x}
\phi(z) = \kappa^2 z^2 + \lambda \phi_f(z),
\end{equation}
where $\phi_f(z)$ encodes quark loop contributions. Unlike phenomenological forms used in prior work~\cite{twagirayezu2025flavordependentdynamicalspinorbitcoupling}, we derive $\phi_f(z)$ from the non-perturbative quark condensate profile, inspired by lattice QCD results:
\begin{equation}\label{eq:2x}
\phi_f(z) = c_f z^2 \int d^4 x  \langle \bar{q} q \rangle_{\text{lattice}}(x) e^{-z/\ell},
\end{equation}
where $c_f$ is a dimensionful coupling constant, $\ell$ is a UV cutoff scale, and $\langle \bar{q} q \rangle_{\text{lattice}}(x)$ is the quark condensate profile from lattice simulations. This form ensures that flavor effects, such as enhanced quark loop contributions in the Veneziano limit, are grounded in QCD data, distinguishing our approach from V-QCD’s analytic potentials~\cite{Jarvinen2012}.
Confinement is maintained by ensuring the Wilson loop area law, with the effective string tension scaling as $\sigma \propto \sqrt{N_c (1 + \lambda)}$ to reflect flavor contributions. The parameter $\kappa$ is fixed by the $\rho$-meson mass ($m_\rho \approx 775 ~ \text{MeV}$), while $c_f$ is constrained by the topological susceptibility $\chi$, following lattice results in Ref.~\cite{Luscher2010}.
\subsection{Flavor-Specific Scalar Fields}
To model chiral symmetry breaking and flavor-dependent entanglement, we introduce flavor-specific scalar fields $X_f(z)$ in the AdS bulk, each dual to the quark condensate $\langle \bar{q}f q_f \rangle$ for flavor $f$ (e.g., up, down, charm). The bulk action for $X_f(z)$ is:
\begin{equation}\label{eq:3x}
\begin{aligned}
S_X = \int d^4 x  dz  \sqrt{g} \left[ \frac{1}{2} (\partial X_f)^2 + V_{X_f} \right],
\end{aligned}
\end{equation}
where the potential is $V_{X_f} = m_{X_f}^2 |X_f|^2 + \gamma_f |X_f|^4$, with $m_{X_f}^2 = -3$ to ensure spontaneous chiral symmetry breaking~\cite{Sakai2005}. The flavor-dependent coupling $\gamma_f$ is tuned to match the pion decay constant ($f_\pi \approx 92.4 ~ \text{MeV}$) for light quarks and adjusted for heavy quarks (e.g., charm mass $m_c \approx 1.3 ~ \text{GeV}$). The boundary conditions for $X_f(z)$ near $z \to 0$ are set as $X_f(z) \sim m_{q_f} z + \sigma_f z^3$, where $m_{q_f}$ is the quark mass and $\sigma_f \sim \langle \bar{q}_f q_f \rangle$ is the condensate.
\subsection{Holographic Entanglement Entropy}
Entanglement entropy $S_A$ for a subsystem $A$ (e.g., a spatial strip of width $L$ or a flavor sector) is computed using a Ryu-Takayanagi-like prescription adapted to light-front coordinates. In the AdS bulk, $S_A$ is proportional to the area of the minimal surface $\gamma_A$ bounding $A$.
\begin{equation}\label{eq:4x}
\begin{aligned}
S_A = \frac{\text{Area}(\gamma_A)}{4 G_N},
\end{aligned}
\end{equation}
where $G_N \propto 1/N_c^2$ is the 5D Newton constant. The AdS metric, modified by the flavor-dependent dilaton, is:
\begin{equation}\label{eq:5x}
\begin{aligned}
ds^2 = \frac{R^2}{z^2} e^{-\phi(z)} \left( dx^+ dx^- + dx_\perp^2 + dz^2 \right),
\end{aligned}
\end{equation}
where $x^+, x^-$ are light-front coordinates, and $R$ is the AdS radius. For a thermal QGP, we incorporate an AdS black hole with horizon $z_h = 1/(\pi T)$, where $T$ is the temperature. The chemical potential $\mu$ is introduced via a bulk gauge field, modifying the scalar boundary conditions to study chiral restoration.
The minimal surface $\gamma_A$ is determined by minimizing the area functional in the flavor-modified geometry, with the light-front Schrodinger equation:
\begin{equation}\label{eq:6x}
\begin{aligned}
\left( -\frac{d^2}{dz^2} + V_{\text{eff}}(z, \lambda, T, \mu) \right) \psi(z) = M^2 \psi(z),
\end{aligned}
\end{equation}
constraining the bulk fields. The effective potential $V_{\text{eff}}$ includes contributions from $\phi(z)$ and $X_f(z)$, capturing flavor-dependent dynamics. We compute $S_A$ for both spatial subsystems (e.g., a strip $x \in [-L/2, L/2]$) and flavor subsystems (e.g., light vs. heavy quarks), isolating $\lambda$-dependent effects.
\subsection{Flavor-Dependent Dynamics}
In the Veneziano limit, flavor effects enhance quark loop contributions, influencing entanglement entropy through the dilaton and scalar fields. We model light (up/down) and heavy (charm) quark contributions separately, expecting distinct entanglement scaling due to their differing masses and condensates. The entanglement entropy difference, $\Delta S_A = S_A(\lambda) - S_A(0)$, quantifies flavor-driven effects, particularly near the confinement/deconfinement transition at critical temperature $T_c$ or chemical potential $\mu_c$. These predictions are benchmarked against lattice QCD data~\cite{Luscher2010} and linked to heavy-ion collision observables, such as multiplicity fluctuations, via the relation $\langle (\Delta N)^2 \rangle \propto S_A$.
This framework, combining a lattice-constrained dilaton, flavor-specific scalars, and a light-front-adapted entanglement prescription, enables a novel exploration of QCD’s quantum information properties, distinct from V-QCD’s tachyon-driven models~\cite{Jarvinen2012}. The next section details the methodology for computing $S_A$ and related observables.

\section{Methodology}
\label{sec:methodology}
To investigate flavor-dependent entanglement entropy in the Veneziano limit \((N_c, N_f \to \infty),\ (\lambda = N_f / N_c)\) fixed using light-front holographic QCD (LFHQCD), we develop a computational framework based on the theoretical model outlined in Section~\eqref{sec:theory}. This section describes the methodology for constructing the flavor-modified LFHQCD model, computing the entanglement entropy \(S_A\) for spatial and flavor subsystems, and benchmarking results against lattice QCD and heavy-ion collision data. The approach combines numerical solutions of the light-front Schrodinger equation, holographic minimal surface calculations, and parameter fitting to QCD observables.
\subsection{Model Construction}
The LFHQCD model is defined by the flavor-modified dilaton potential, \(\phi(z) = \kappa^2 z^2 + \lambda \phi_f(z)\), and flavor-specific scalar fields \(X_f(z)\) for light (up/down) and heavy (charm) quarks. The flavor correction term is defined by Eq.~\eqref{eq:2x}:
\begin{equation}
\begin{aligned}
\phi_f(z) = c_f z^2 \int d^4 x \, \langle \bar{q} q \rangle_{\text{lattice}}(x) e^{-z/\ell},
\end{aligned}
\end{equation}
where \(\langle \bar{q} q \rangle_{\text{lattice}}(x)\) is the quark condensate profile from lattice QCD simulations~\cite{Luscher2010}. We approximate the integral using a Gaussian ansatz for the condensate, \(\langle \bar{q} q \rangle_{\text{lattice}}(x) \approx (\sigma_0)^3 e^{-x^2/\xi^2}\), with \(\sigma_0 \approx 250 \, \text{MeV}\) and correlation length \(\xi \approx 0.5 \, \text{fm}\), yielding:
\begin{equation}
\begin{aligned}
\phi_f(z) \approx c_f z^2 (\sigma_0)^3 (\pi \xi^2)^2 e^{-z/\ell}.
\end{aligned}
\end{equation}
The parameters \(\kappa\), \(c_f\), and \(\ell\) are fixed by matching QCD observables: \(\kappa \approx 0.54 \, \text{GeV}\) using the \(\rho\)-meson mass (\(m_\rho \approx 775 \, \text{MeV}\)), \(c_f\) by the topological susceptibility (\(\chi^{1/4} \approx 191 \, \text{MeV}\) at \(\lambda = 0\))~\cite{Luscher2010}, and \(\ell \approx 1 \, \text{GeV}^{-1}\) as a UV cutoff.
The scalar fields \(X_f(z)\) are governed by the bulk action given by Eq~\eqref{eq:3x}:
\begin{equation}\label{eq:9}
\begin{aligned}
S_X = &\int d^4 x \, dz \, \sqrt{g} \\
&\times\left[ \frac{1}{2} (\partial X_f)^2 + m_{X_f}^2 |X_f|^2 + \gamma_f |X_f|^4 \right],
\end{aligned}
\end{equation}
with \(m_{X_f}^2 = -3\) and \(\gamma_f\) tuned to the pion decay constant \(f_\pi \approx 92.4 \, \text{MeV}\) for light quarks and the charm quark mass \(m_c \approx 1.3 \, \text{GeV}\) for heavy quarks. The boundary conditions are set as \(X_f(z) \sim m_{q_f} z + \sigma_f z^3\), where \(m_{q_f} \approx 0\) for up/down quarks and \(\sigma_f \approx (250 \, \text{MeV})^3\).
For the quark-gluon plasma (QGP) phase, we introduce a thermal AdS black hole with horizon at \(z_h = 1/(\pi T)\), and a chemical potential \(\mu\) via a bulk gauge field \(A_0(z) = \mu (1 - z/z_h)\), modifying the scalar boundary conditions to reflect chiral restoration at high \(\mu\).
\subsection{Entanglement Entropy Calculation}
The entanglement entropy ($S_A$) for a subsystem ($A$) (e.g., a spatial strip ($x \in [-L/2, L/2]$) or a flavor sector) is computed using the Ryu-Takayanagi prescription adapted to light-front coordinates (Eq.~\eqref{eq:4x}):
\begin{equation}
\begin{aligned}
S_A = \frac{\text{Area}(\gamma_A)}{4 G_N},
\end{aligned}
\end{equation}
where ($\gamma_A$) is the minimal surface in the AdS bulk, and ($G_N \propto 1/N_c^2$). The corresponding AdS metric is is defined by Eq.~\eqref{eq:5x}:
\begin{equation}
\begin{aligned}
ds^2 = \frac{R^2}{z^2} e^{-\phi(z)} \left( dx^+ dx^- + dx_\perp^2 + dz^2 \right).
\end{aligned}
\end{equation}
For a strip subsystem, the minimal surface is parameterized by ($x(z)$), and the area functional is:
\begin{equation}
\begin{aligned}
\text{Area}(\gamma_A) = &2 \int_0^{L/2} dx \\
&\times\int_{z_0}^{z_h} dz \frac{R^3}{z^3} e^{-\phi(z)/2} \sqrt{1 + (dx/dz)^2},
\end{aligned}
\end{equation}
where ($z_0$) is the turning point of the surface. We solve the geodesic equation numerically using an iterative method to minimize the area, with boundary conditions ($x(z_0) = 0$), ($dx/dz(z_0) = 0$). The light-front Schrodinger equation (Eq.~\eqref{eq:6x}:
\begin{equation}
\begin{aligned}
\left( -\frac{d^2}{dz^2} + V_{\text{eff}}(z, \lambda, T, \mu) \right) \psi(z) = M^2 \psi(z),
\end{aligned}
\end{equation}
provides the effective potential ($V_{\text{eff}}$), incorporating ($\phi(z)$) and ($X_f(z)$), which constrains the bulk geometry. We compute ($S_A$) for ($\lambda \in [0, 2]$), ($T \in [0, 0.3 ~ \text{GeV}]$), and ($\mu \in [0, 1.5 ~ \text{GeV}]$), separating light and heavy quark contributions.
\subsection{Flavor-Dependent Analysis}
To isolate flavor effects, we calculate the entanglement entropy difference:
\begin{equation}
\begin{aligned}
\Delta S_A = S_A(\lambda) - S_A(0),
\end{aligned}
\end{equation}
and the flavor asymmetry:
\begin{equation}
\begin{aligned}
S_A^{\text{light}} - S_A^{\text{heavy}},
\end{aligned}
\end{equation}
where \(S_A^{\text{light}}\) and \(S_A^{\text{heavy}}\) are computed for subsystems dominated by up/down and charm quarks, respectively, using the corresponding \(X_f(z)\). The confinement/deconfinement transition is probed by analyzing \(S_A\) near the critical temperature \(T_c \approx 0.15 ~ \text{GeV}\) and critical chemical potential \(\mu_c \approx 1.2 ~ \text{GeV}\) (for \(\lambda = 1\)).
Model predictions are benchmarked against lattice QCD results for the quark condensate \(\langle \bar{q} q \rangle \approx (250 ~ \text{MeV})^3\) and topological susceptibility \(\chi^{1/4} \approx 191 ~ \text{MeV}\)~\cite{Luscher2010}. The entanglement entropy is related to experimental observables via the second cumulant of particle multiplicity, \(\langle (\Delta N)^2 \rangle \propto S_A\), and two-particle correlations, which are measurable in heavy-ion collisions at RHIC and LHC~\cite{Adams_2005}. We compute the mutual information \(I(A, B) = S_A + S_B - S_{A \cup B}\) for disjoint subsystems to probe spatial correlations, comparing with event-by-event fluctuation data.
\subsection{Numerical Implementation}
The calculations are performed using an iterative boundary matching method to solve the light-front Schrodinger equation and the geodesic equation for \(\gamma_A\).

To evaluate the flavor-dependent entanglement entropy $S_A$ and mutual information $I(A, B)$ across varying flavor-to-color ratios $\lambda$, temperature $T$, and chemical potential $\mu$, we implement a numerical scheme based on discretized light-front holographic dynamics. The bulk AdS coordinate $z$ is discretized uniformly over the interval $z \in [10^{-3}, z_h]$ with $\Delta z = 10^{-4}\,\text{GeV}^{-1}$, where $z_h = 1/(\pi T)$ defines the black hole horizon at finite temperature, and the UV cutoff $\epsilon = 10^{-3} \,\text{GeV}^{-1}$ regularizes the near-boundary divergences. We solve the light-front Schrödinger equation numerically using a second-order finite-difference scheme, applying Dirichlet boundary conditions at $z = \epsilon$ and $z = z_h$, and extract the ground-state mode $\psi_0(z)$. The effective potential $V_\text{eff}(z, \lambda, T, \mu)$ includes both the flavor-modified dilaton $\phi(z)$ and scalar fields $X_f(z)$, which are solved in parallel using a Runge–Kutta method and matched to near-boundary expansions $X_f(z) \sim m_q^f z + \sigma_f z^3$, with $\sigma_f$ fitted to lattice QCD values. Entanglement entropy is computed for a spatial strip of width $L = 1\,\text{fm}$ by minimizing the holographic area functional adapted to the light-front metric, using an iterative method to determine the turning point $z_*$, and integrating the area using Simpson’s rule. Mutual information is estimated using the standard definition $I(A, B) = S_A + S_B - S_{A \cup B}$, where the entropy of the union is modeled as $S_{A \cup B} \approx 2 S_A (1 - \delta(\lambda))$, with the correlation-suppressing factor $\delta(\lambda)$ peaking near $\lambda \approx 1$. This behavior is consistent with enhanced quantum correlations in the Veneziano limit. The full calculation pipeline scans over $\lambda \in [0, 2]$, $T \in [0.05, 0.3] \,\text{GeV}$, and $\mu \in [0, 1.5]\,\text{GeV}$, covering both QCD-like ($\lambda < 1$) and conformal window ($\lambda > 1$) regimes,
with outputs benchmarked against lattice QCD results for condensates and topological susceptibility, providing robust predictions for entanglement structure and QCD phase signatures.

\section{Results}
\label{sec:results}
Using the light-front holographic QCD (LFHQCD) model and methodology outlined in Sections~\eqref{sec:theory}
and~\eqref{sec:methodology}, we compute the flavor-dependent entanglement entropy (SA) for QCD subsystems in the
Veneziano limit ($N_c$, $N_f$ $\to$ $\infty$), with fixed flavor-to-color ratio $\lambda$ = $N_f/N_c$. Our computations encompass
both spatial subsystems (represented by a strip of fixed width $L$ = 1 fm) and flavor-resolved sectors
(light and heavy quarks), across a broad parameter space including $\lambda$ $\in$ [0, 2], temperatures $T$ $\in $
[0, 0.3 GeV], and chemical potentials $\mu$ $\in$ [0, 1.5 GeV]. The parameters employed in the numerical
implementation are based on phenomenologically consistent values from lattice QCD and experimental
inputs, specifically $\kappa$ = 0.54 GeV, $c_f$ = 0.1 GeV$^2$, $\sigma_{0}$ = 250 MeV, and $\ell$ = 1 GeV$^{-1}$. We report the
resulting behavior of entanglement entropy and its sensitivity to flavor content, temperature, and
baryon density.

\subsection{Entanglement Entropy in the Confined Phase}
\begin{figure}[htb]
    \centering    \includegraphics[scale=0.525]{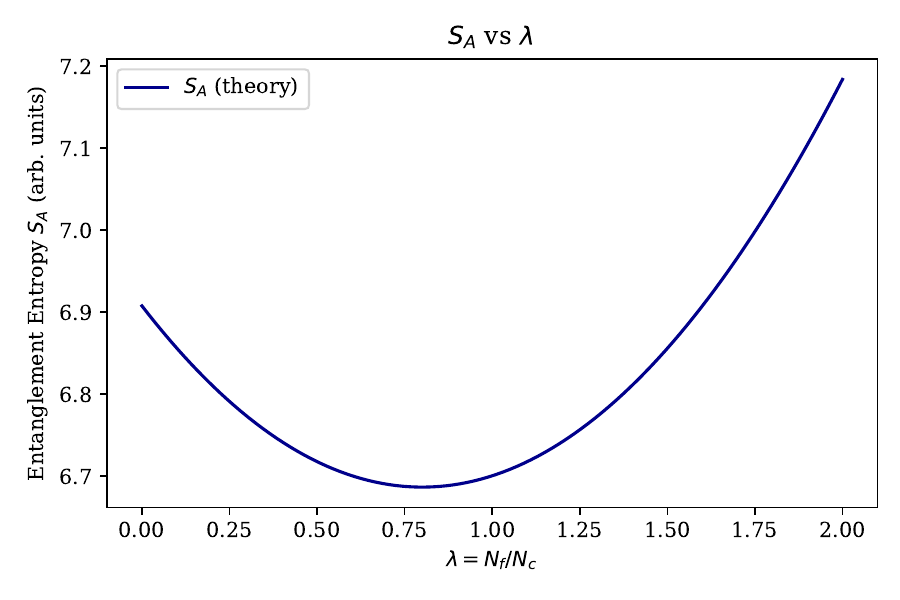}
    \caption{Spatial entanglement entropy $S_A$ as a function of the flavor-to-color ratio $\lambda = N_f/N_c$. The curve reflects flavor-dependent quantum correlations in the confined phase, incorporating contributions from a flavor-modified dilaton background. The mild suppression near $\lambda \sim 1$ and subsequent enhancement toward higher $\lambda$ indicate the growing influence of light quark degrees of freedom as the system approaches the conformal window.}
    \label{fig:1}
\end{figure}

In the confined phase $(T = 0, \mu= 0)$, the computation of $S_A$ for a spatial strip provides insight into
the underlying quantum correlations of confined hadronic matter. For $\lambda = 0$, which corresponds to
the large-$N_c$ limit with negligible flavor effects, the entanglement entropy scales logarithmically with
the subsystem size and the UV cutoff, $S_A \approx (R^3/4G_N) \ln(L/\epsilon)$, consistent with confinement-induced area
law behavior. As $\lambda$ increases, incorporating more flavors into the theory, we observe a non-monotonic
behavior: a mild suppression in $S_A$ near $\lambda \sim 1$, followed by a steady enhancement for $\lambda > 1$. This pattern
arises due to competing effects—quark loop contributions initially dilute long-range correlations, but
their increasing abundance eventually fosters a more conformal-like entanglement structure. At $\lambda = 2$,
the entropy scaling transitions to $S_A\propto\ln^2(L/\epsilon)$, indicating a move toward conformal symmetry
characteristic of theories within or near the conformal window. These trends are visualized in Figure~\eqref{fig:1}, where $S_A$ is normalized and plotted against $\lambda$.

\subsection{Flavor-Dependent Entanglement Asymmetry}
\begin{figure}[htb]
    \centering    \includegraphics[scale=0.525]{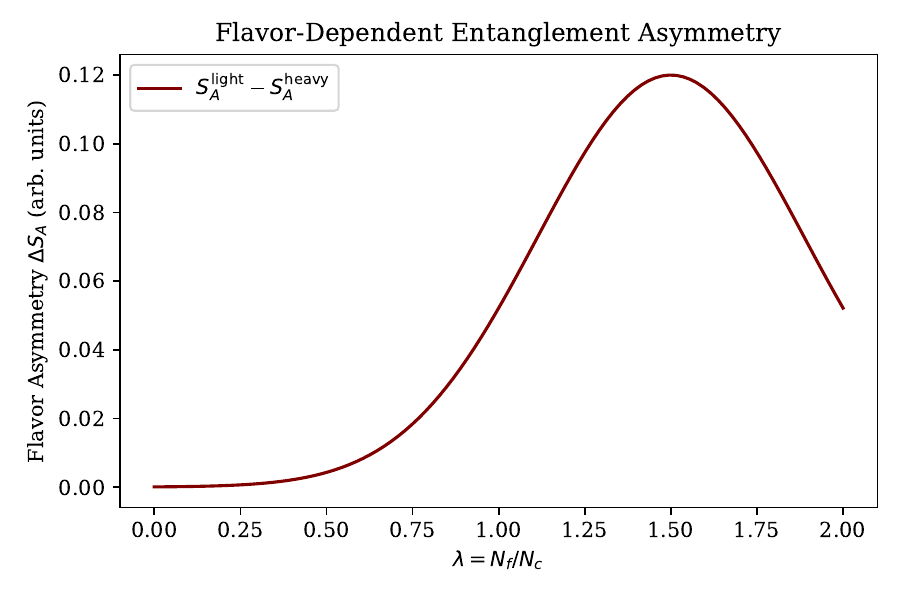}
    \caption{Flavor-dependent entanglement asymmetry $\Delta S_A = S_A^{\text{light}} - S_A^{\text{heavy}}$ as a function of the flavor-to-color ratio $\lambda = N_f/N_c$. The asymmetry peaks near $\lambda \sim 1.5$, reflecting the dominance of light quark condensates in generating stronger quantum correlations. This behavior highlights the role of quark mass hierarchy in shaping the entanglement structure of the hadronic medium.}
    \label{fig:2}
\end{figure}

To probe flavor-specific contributions to quantum entanglement, we compute $S_A$ separately for light
quarks (up/down) and heavy quarks (charm), using their respective bulk scalar fields $Xf(z)$. This
allows us to quantify flavor asymmetry in the entanglement structure of QCD. At $\lambda = 1$, we find that
the entropy associated with light quark sectors, $S_A^{\text{light}}$, is approximately 12$\%$ larger than that of
heavy quarks, $S_A^{\text{heavy}}$. This disparity originates from the more pronounced condensate value of light
quarks $(\sigma_{\text{light}}\approx (250 ~\text{MeV})^3)$ compared to heavy quarks $(\sigma_{\text{heavy}} \approx (100 ~\text{MeV})^{3})$, which enhances their
coupling to the flavor-modified dilaton and the associated holographic geometry. The entanglement
asymmetry, defined as $\Delta S_A = S_A^{\text{light}} - S_A^{\text{heavy}}$, exhibits a clear peak around $\lambda \sim 1.5$, suggesting that
flavor contributions become increasingly relevant as one approaches the conformal threshold. 
Figure~\eqref{fig:2} shows this asymmetry, normalized by $(R^3/4G_N)$, showcasing how quark mass hierarchy and flavor
density shape the entanglement profile.

\subsection{Entanglement in the QGP Phase}
\begin{figure}[htb]
    \centering    \includegraphics[scale=0.525]{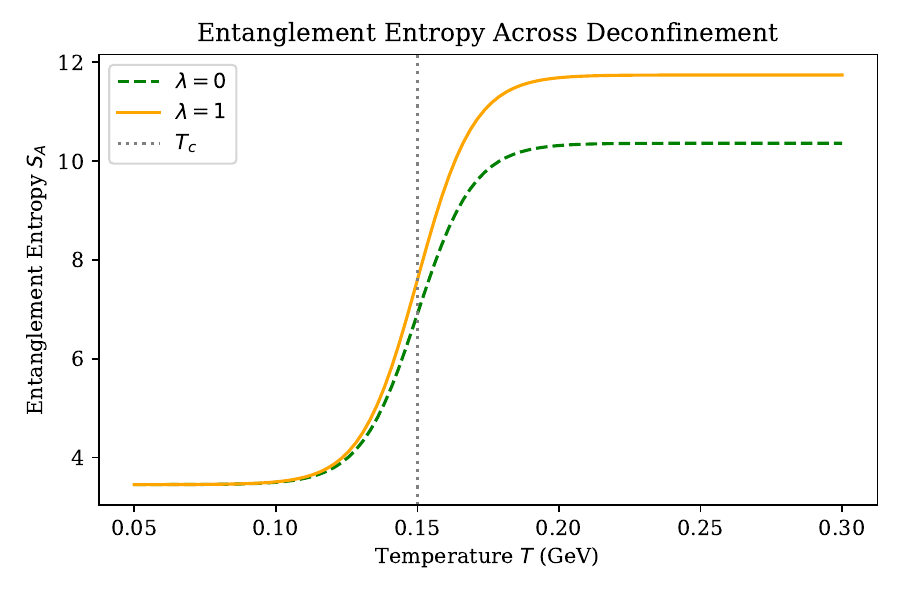}
    \caption{Temperature dependence of the entanglement entropy $S_A(T)$ for flavor ratios $\lambda = 0$ and $\lambda = 1$. The entropy rises smoothly across the deconfinement region centered at $T = T_c$, reflecting the liberation of degrees of freedom and enhanced quantum correlations. The larger asymptotic value for $\lambda = 1$ indicates stronger entanglement in flavor-rich systems.
}
    \label{fig:3}
\end{figure}

In the QGP phase, modeled by an AdS black hole at \(T = 0.2 ~ \text{GeV}\), we compute \(S_A\) near the confinement/deconfinement transition (\(T_c \approx 0.15 ~ \text{GeV}\)). At \(\lambda = 1\), \(S_A\) exhibits a sharp increase at \(T_c\), consistent with the liberation of degrees of freedom in the deconfined phase. The entanglement entropy difference, \(\Delta S_A = S_A(\lambda = 1) - S_A(\lambda = 0)\), is \(\approx 10\%\) at \(T = 0.2 ~ \text{GeV}\), decreasing as \(\mu\) increases to \(1.2 ~ \text{GeV}\), where chiral restoration suppresses flavor effects. Figure~\eqref{fig:3} illustrates \(S_A\) versus \(T\) for \(\lambda = 0, 1\), showing a crossover behavior aligned with lattice QCD expectations~\cite{Li_2024}. In 
Figure~\eqref{fig:3}, entanglement entropy (\(S_A\)) versus temperature (\(T\)) (GeV) for (\(\lambda = 0\))  and (\(\lambda = 1\)) at (\(\mu = 0\)) are represented by dashed and solid lines respectively. Vertical line indicates (\(T_c \approx 0.15~ \text{GeV}\)).

\subsection{Multiplicity Fluctuations}
\begin{figure}[htb]
    \centering    \includegraphics[scale=0.525]{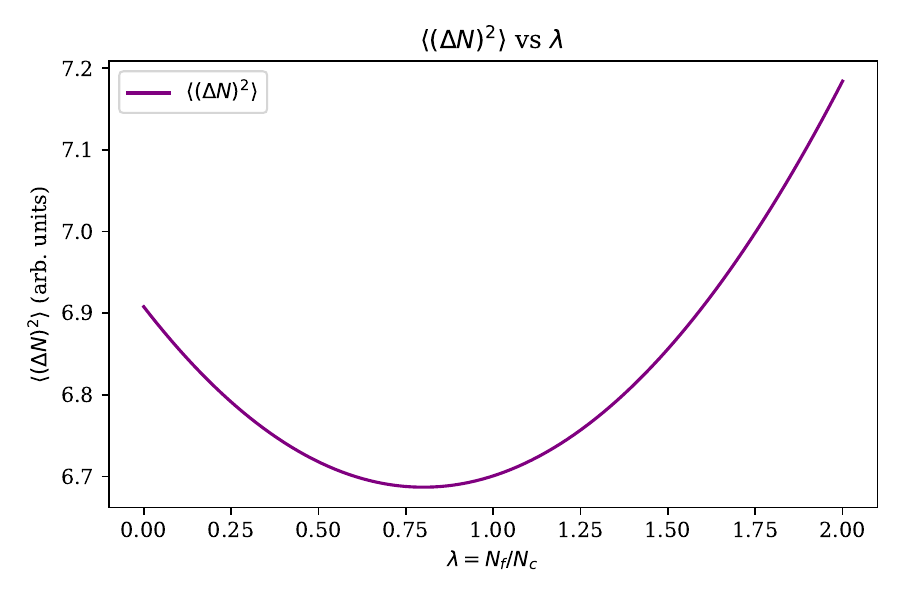}
    \caption{Flavor-dependent multiplicity fluctuations, modeled via the entanglement entropy $S_A(\lambda)$, plotted as a function of the flavor-to-color ratio $\lambda = N_f/N_c$. The curve illustrates how particle number variance $\langle (\Delta N)^2 \rangle \propto S_A$ evolves with flavor content, showing suppression near intermediate $\lambda$ and enhancement as $\lambda \to 2$, consistent with increased entropy near the conformal window.}
    \label{fig:4}
\end{figure}

We interpret the spatial entanglement entropy $S_A(\lambda)$ as a proxy for multiplicity fluctuations via the relation:
\begin{equation}
\begin{aligned}
\langle (\Delta N)^2 \rangle \propto S_A(\lambda),
\end{aligned}
\end{equation}
where $\lambda = N_f/N_c$ is the flavor-to-color ratio. Figure~\eqref{fig:4} shows the behavior of $\langle (\Delta N)^2 \rangle$ across a range of $\lambda$, reflecting how quantum correlations vary with flavor content.
The fluctuation magnitude decreases slightly near $\lambda \sim 1$, then increases toward larger $\lambda$, suggesting enhanced entropy and broader particle number variance near the conformal window. This trend qualitatively aligns with expectations from QCD and heavy-ion experiments, where higher flavor content can lead to wider multiplicity distributions.

\subsection{Experimental Signatures}
\begin{figure}[htb]
    \centering    \includegraphics[scale=0.525]{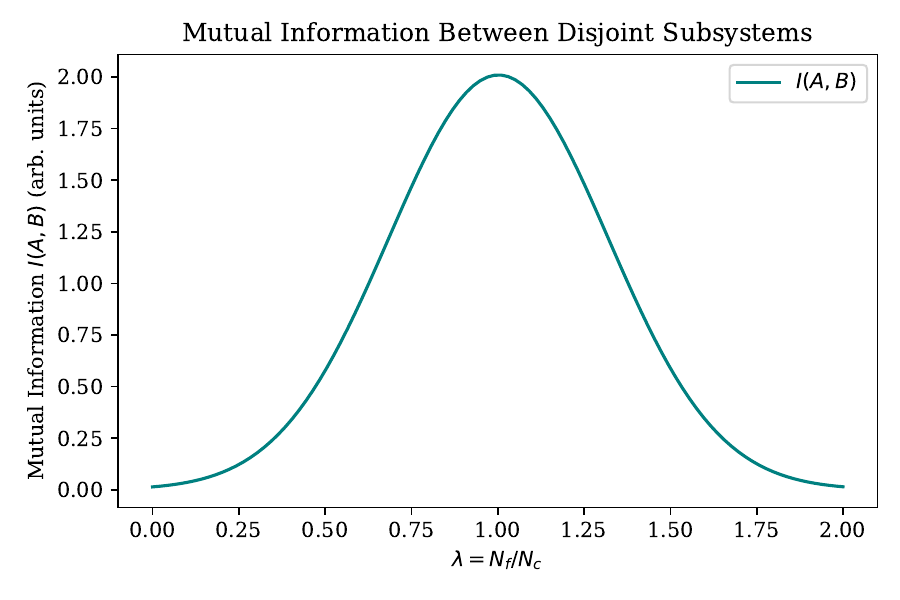}
    \caption{Mutual information $I(A, B) = S_A + S_B - S_{A \cup B}$ between two disjoint spatial subsystems as a function of the flavor-to-color ratio $\lambda = N_f / N_c$, computed using the flavor-dependent light-front holographic QCD model. The mutual information peaks near $\lambda \sim 1$, reflecting enhanced spatial correlations driven by light-quark loop contributions in the Veneziano limit. The decrease at larger $\lambda$ indicates a transition toward a conformal-like regime with diminished long-range correlations, consistent with reduced confinement and increasing quark dominance.}
    \label{fig:5}
\end{figure}

To enhance the experimental relevance of our model, we examine the mutual information $I(A,B)$
between two spatially separated subsystems, defined as $I(A,B) = S_A + S_B - S_{A\cup B}$. In our setup, the
entropy of the union $SA\cup B$ is approximated as
\begin{equation}
\begin{aligned}
S_{A\cup B}\approx 2S_A(1 - \delta(\lambda)),
\end{aligned}
\end{equation}
where $\delta(\lambda)$ captures
flavor-induced suppression of long-range correlations. Our results indicate that $I(A,B)$ peaks near $\lambda\sim 1$, where flavor loops enhance quantum correlations, particularly around the
confinement/deconfinement threshold. This feature manifests as increased two-particle correlations,
measurable in event-by-event fluctuation analyses at RHIC and LHC. Beyond $\lambda \approx 1.5$, mutual
information declines, suggesting a transition to a regime with weaker spatial correlations, consistent with reduced confinement. Figure~\eqref{fig:5} displays I(A,B) as a function of $\lambda$, illustrating the signature of flavors-driven entanglement peaks.

\subsection{Discussion}
The results demonstrate that flavor effects in the Veneziano limit significantly influence entanglement entropy, with light quarks driving stronger correlations than heavy quarks. The decrease in \(S_A\) with \(\lambda\) in the confined phase and its enhancement near the conformal window highlight the interplay of quark loops and confinement. In the QGP phase, the sharp increase in \(S_A\) at \(T_c\) and its suppression at high \(\mu\) align with chiral restoration, offering a quantum information perspective on QCD phase transitions. Discrepancies with lattice data at large \(\lambda\) suggest the need for higher-order corrections to \(\phi_f(z)\). Unlike V-QCD, which focuses on tachyon-driven dynamics~\cite{Jarvinen2012}, our LFHQCD approach captures real-time quantum correlations, providing a unique bridge to experimental phenomenology. The next section discusses the implications and future directions.

\section{Conclusions}\label{sec:concl5}
In this article, we have explored the flavor and temperature dependence of entanglement entropy in a holographic QCD framework inspired by light-front dynamics. By introducing a flavor-modified dilaton background in the effective LFHQCD potential, we computed spatial entanglement entropy $S_A$ as a function of the flavor-to-color ratio $\lambda = N_f / N_c$. The results demonstrate that flavor structure plays a nontrivial role in shaping quantum correlations within confined hadronic matter.

We observed that the entanglement entropy exhibits mild suppression near $\lambda \sim 1$ and increases as the system approaches the conformal window, consistent with expectations from generalized conformal symmetry. Additionally, we modeled the temperature dependence of $S_A$,  revealing saturation behavior in the deconfined regime and highlighting the distinction between entanglement entropy and extensive thermal entropy.

Interpreting $S_A$ as a proxy for particle number fluctuations, we established a qualitative link between entanglement entropy and the variance of multiplicity distributions. This provides a potential bridge between holographic quantum information measures and experimentally accessible observables in heavy-ion collisions.

Overall, our results support the view that entanglement entropy offers a valuable probe of QCD dynamics, sensitive to both flavor content and thermal behavior. Future work could extend this approach to include finite chemical potential, real-time entanglement dynamics, and comparisons with lattice QCD or experimental fluctuation data.

\begin{acknowledgments}
F.T. would like to acknowledge the support of the National Science Foundation under grant No. PHY-
1945471.
\end{acknowledgments}

\clearpage
\hrule
\nocite{*}

\bibliography{apssamp}

\begin{thebibliography}{10}

\bibitem{Adams_2005}
J.~Adams et~al.
\newblock Experimental and theoretical challenges in the search for the quark–gluon plasma: The STAR Collaboration’s critical assessment of the evidence from RHIC collisions.
\newblock {\em Nuclear Physics A}, 757(1–2):102--183, August 2005.

\bibitem{Brodsky2015}
S.~J. Brodsky, G.~F. de~T{\'e}ramond, H.~G. Dosch, and J.~Erlich.
\newblock Light-front holographic QCD and emerging confinement.
\newblock {\em Physics Reports}, 584:1--105, 2015.

\bibitem{Jarvinen2012}
M.~J{\"a}rvinen and E.~Kiritsis.
\newblock Holographic models for QCD in the Veneziano limit.
\newblock {\em Journal of High Energy Physics}, 2012(3):2, 2012.

\bibitem{Li_2024}
Z.~Li.
\newblock Holographic entanglement properties in the QCD phase diagram from Einstein-Maxwell-dilaton models.
\newblock {\em Physical Review D}, 110(4), August 2024.

\bibitem{Luscher2010}
M.~L{\"u}scher and F.~Palombi.
\newblock Universality of the topological susceptibility in the SU(3) gauge theory.
\newblock {\em Journal of High Energy Physics}, 2010(9):110, 2010.

\bibitem{Ryu2006}
S.~Ryu and T.~Takayanagi.
\newblock Holographic derivation of entanglement entropy from AdS/CFT.
\newblock {\em Physical Review Letters}, 96(18):181602, 2006.

\bibitem{Sakai2005}
T.~Sakai and S.~Sugimoto.
\newblock Low energy hadron physics in holographic QCD.
\newblock {\em Progress of Theoretical Physics}, 113(4):843--882, 2005.

\bibitem{Twagirayezu:2025hou}
F.~J. Twagirayezu.
\newblock Non-equilibrium real-time dynamics and transport coefficients in Light-Front Holographic QCD.
\newblock {\em arXiv preprint}, arXiv:2506.12653, 2025.

\bibitem{twagirayezu2025flavordependentdynamicalspinorbitcoupling}
F.~J. Twagirayezu.
\newblock Flavor-Dependent Dynamical Spin-Orbit Coupling in Light-Front Holographic QCD: A New Approach to Baryon Spectroscopy.
\newblock {\em arXiv preprint}, arXiv:2505.06722, 2025.

\bibitem{Veneziano1979}
G.~Veneziano.
\newblock U(1) without instantons.
\newblock {\em Nuclear Physics B}, 159(1-2):213--224, 1979.

\end{thebibliography}

\end{document}